\documentclass[twocolumn,aps,10pt,prl,showpacs,showkeys,]{revtex4}
\usepackage[dvips]{graphicx}

\begin{document}
\title{Crystal field, spin-orbit coupling and magnetism in a ferromagnet YTiO$_{3}$$^\spadesuit$}
\author{R. J. Radwanski}
\affiliation{Center of Solid State Physics, S$^{nt}$Filip 5,
31-150 Krakow, Poland,
\\
Institute of Physics, Pedagogical University, 30-084 Krakow,
Poland} \homepage{http://www.css-physics.edu.pl}
\email{sfradwan@cyf-kr.edu.pl}
\author{Z. Ropka}
\affiliation{Center of Solid State Physics, S$^{nt}$Filip 5,
31-150 Krakow, Poland}

\begin{abstract}
Magnetic properties of stechiometric YTiO$_{3}$ has been
calculated within the single-ion-based paradigm taking into
account the low-symmetry crystal field and the intra-atomic
spin-orbit coupling of the Ti$^{3+}$ ion. Despite of the very
simplified approach the calculations reproduce perfectly the
value of the magnetic moment and its direction as well as
temperature dependence of the magnetic susceptibility $\chi (T)$.
It turns out that the spin-orbit coupling is fundamentally
important for 3d magnetism  and magnetic properties are
determined by lattice distortions.

\end{abstract}
\maketitle \vspace {-0.3 cm}

\section {Introduction}
\vspace {-0.6 cm} YTiO$_3$ is a unique 3$d$ ferromagnet
\cite{1,2,3} - the most of oxides are antiferromagnetic. In
combination with a rather simple perovskite structure and
Ti$^{3+}$ ions expected to have one electron in the incomplete
3$d$ shell, YTiO$_{3}$ is regarded to be very good examplary
system for studying basic interactions in 3$d$ oxides. Despite
this simplicity its properties are not understood yet. In fact,
there is going on at present a hot debate on description of its
magnetic and electronic properties
\cite{4,5,6,7,8,9,10,11,12,13,14,15,16,17,18,19,20}. A time when
such system was treated as a S=1/2 system i.e. with the spin-only
magnetism and the fully quenched orbital magnetism is already
gone. Also the simplest versions of the band picture turned out
to be completely useless giving already at the start disagreement
with experiment predicting a metallic ground state whereas
YTiO$_{3}$ is in the reality a good insulator.

YTiO$_{3}$, when stoichiometric, is very good insulator. Its
resistivity at room temperature amounts to 5$\cdot$10$^{-2}$
$\Omega$cm \cite{11}.  The resistivity rapidly grows up with
decreasing temperature. It is ferromagnetic below T$_{c}$ of
30-35 K \cite{3,4}. The macroscopic magnetisation, if
recalculated per the formula unit, points to a moment of 0.84
$\mu_{B}$ \cite{3}. The paramagnetic susceptibility has been
found to follow the Curie-Weiss law with $\theta$ = 39 K, if
substracted a diamagnetic and temperature independent orbital
contributions of a total size of 0.351$\cdot$10$^{-3}$ emu/mol.

The aim of this paper is to present results of single-ion based
calculations of properties of YTiO$_{3}$. Our single-ion results
seem to be quite remarkable. We took into account low-symmetry
off-octahedral crystal-field (CEF) interactions and the
intra-atomic spin-orbit (s-o) coupling, that turn out to be of the
comparable strength. \vspace {-1.0 cm}

\section {Theoretical outline}
\vspace {-0.5 cm} We consider exactly stechiometric YTiO$_{3}$.
From this and the insulating ground state we infer that all Ti
ions are in the trivalent state according to the charge
distribution Y$^{3+}$Ti$^{3+}$O$_{3}^{2-}$. The relevant charge
transfer occurs during the formation of the compound. In the
perovskite-based structure of YTiO$_{3}$ the Ti$^{3+}$ ion is
surrounded by six oxygen ions forming distorted octahedron. There
are still some Ti ions at the surface, for instance, with a
reduced symmetry, but they are generally neglected, because we are
interested in intrinsic properties of YTiO$_{3}$. The local
octahedra in YTiO$_{3}$ are tilted and rotated, what causes the
need for consideration a larger elementary cell, with four Ti ions
instead of one as in the simple perovskite structure. Thanks it
other three Ti ions get a crystallographic freedom. As a result of
rotations, tilts and other atom displacements there are four
short Ti-O bonds and two long Ti-O bonds.

The orthorhombic lattice in the Pbnm structure results from that
of an ideal cubic perovskite by setting $a_{o}$ =$\surd{2}$
$a_{c}$, $b_{o}$ =$\surd{2}$ $a_{c}$ and $c_{o}$ =2a$_{c}$, where
$a_{c}$ denotes the lattice parameter of the simple cubic
perovskite. $a_{c}$ is of order of 400 pm.

In the $Pnma$ structure used by some authors the doubling occurs
along the $b$ direction. The $c$, $a$, and $b$ axis in the $Pnma$
structure becomes the $a$, $b$ and $c$ axis in the $Pbnm$
structure, respectively. Independently of the used $Pbnm$ or
$Pnma$ space group the derivation of the local surroundings must
be, of course, the same.

Here we use the $Pbnm$ space group. Then the easy magnetic axis is
the $c$ axis, \cite {3}. The lattice parameters at T = 293 K,
according to Ref. 2, cited by \cite {5,15}, are: $a_{o}$ =531.6
pm, $b_{o}$=567.9 pm and $c_{o}$ = 761.1 pm. These parameters
have been confirmed by detailed structural measurements of Loa
{\it et al.} \cite {19}.

Cwik {\it et al.} \cite {5} have derived the respective bonds:
207.7 pm (Ti-O(b)), 201.6 pm (Ti-O(a)) and the apical bond 202.3
pm (Ti-O(c)). Loa {\it et al.} \cite {19} have measured influence
of the external pressure on these bonds. From these
crystallographic studies we get an input to our theoretical
considerations that the lattice surroundings of the Ti ion in
YTiO$_{3}$ is predominantly octahedral with a slight orthorhombic
distortion.

\section {Results and discussion}
\vspace {-0.5 cm} The electronic structure of the Ti$^{3+}$ ion
with one d-electron under the action of the crystal-field
interactions H$_{CF}$ and in the presence of the intra-atomic
spin-orbit coupling H$_{s-o}$ we calculated with the use of the
single-ion-like Hamiltonian often used in description of 3$d$
impurity states in the Electron Paramagnetic Resonance \cite
{22,23,24}, which we accept also for a solid, where 3$d$ atom is
the full part of the crystallographic structure:
\begin{center}
$H_d=H_{CF}+H_{s-o}=B_4(O_4^0+5O_4^4)+\lambda _{s-o}L\cdot
S+B_2^0O_2^0+\mu _B(L+$g$_eS)\cdot B_{ext}. (1)$
\end{center}
The calculated electronic structure is presented in Fig. 1. The
crystal field has been divided into the cubic part, usually
dominant in case of compounds containing 3$d$ ions, and the
off-octahedral distortion written by the second-order leading
term B$_2^0$O$_2^0$. The last term, Zeeman term, allows
calculations of the influence of the external magnetic field.
$g_e$ amounts to 2.0023. The Zeeman term is necessary for
calculations, for instance, of the paramagnetic susceptibility -
in fact the paramagnetic susceptibility is customarily calculated
[25] as the magnetization in an external field of, say, 0.1 T
applied along different crystallographic directions.
\begin{figure}[t]
\begin{center}
\includegraphics[width = 6.1 cm]{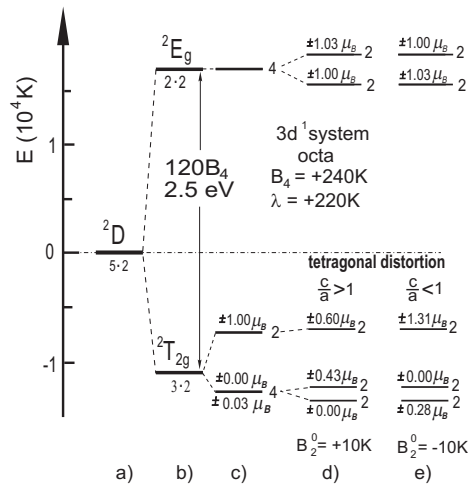}
\end{center} \vspace {-0.7 cm}
\caption{The calculated fine electronic structure of the 3$
d^{1}$ electronic system (Ti$^{3+}$, V$^{4+}$ ions) in the
paramagnetic state under the action of the crystal field and
spin-orbit interactions: a) the 10-fold degenerated $^2$D term
realized in the absence of the CEF and the s-o interactions; b)
the splitting of the $^2$D term by the octahedral CEF surrounding B$_4$%
=+240 K ($\lambda _{s-o}$ =0) yielding the $^2$T$_{2g}$ cubic
subterm as the ground state; c) the splitting of the lowest
$^2$T$_{2g}$ cubic subterm by the combined octahedral CEF and
spin-orbit interactions (B$_4$= +240 K and $\lambda _{s-o}$= +220
K); the degeneracy and the associated magnetic moments are shown;
d) the splitting due to the elongated tetragonal off-octahedral
distortion of B$_2^0$= +10 K (c/a$>$1); e) the splitting due to
the compressing tetragonal distortion of B$_2^0$= -10 K (c/a$<$1,
apical oxygens become closer). Figs c, d and e are not to the
left hand energy scale - the splitting of the three lowest states
on Figs c, d and e amounts to 333 K, 368 K and 372 K,
respectively.\vspace {-0.6 cm}}
\end{figure}
\begin{figure}\vspace {-0.8 cm}
\begin{center}
\includegraphics[width = 6.2 cm]{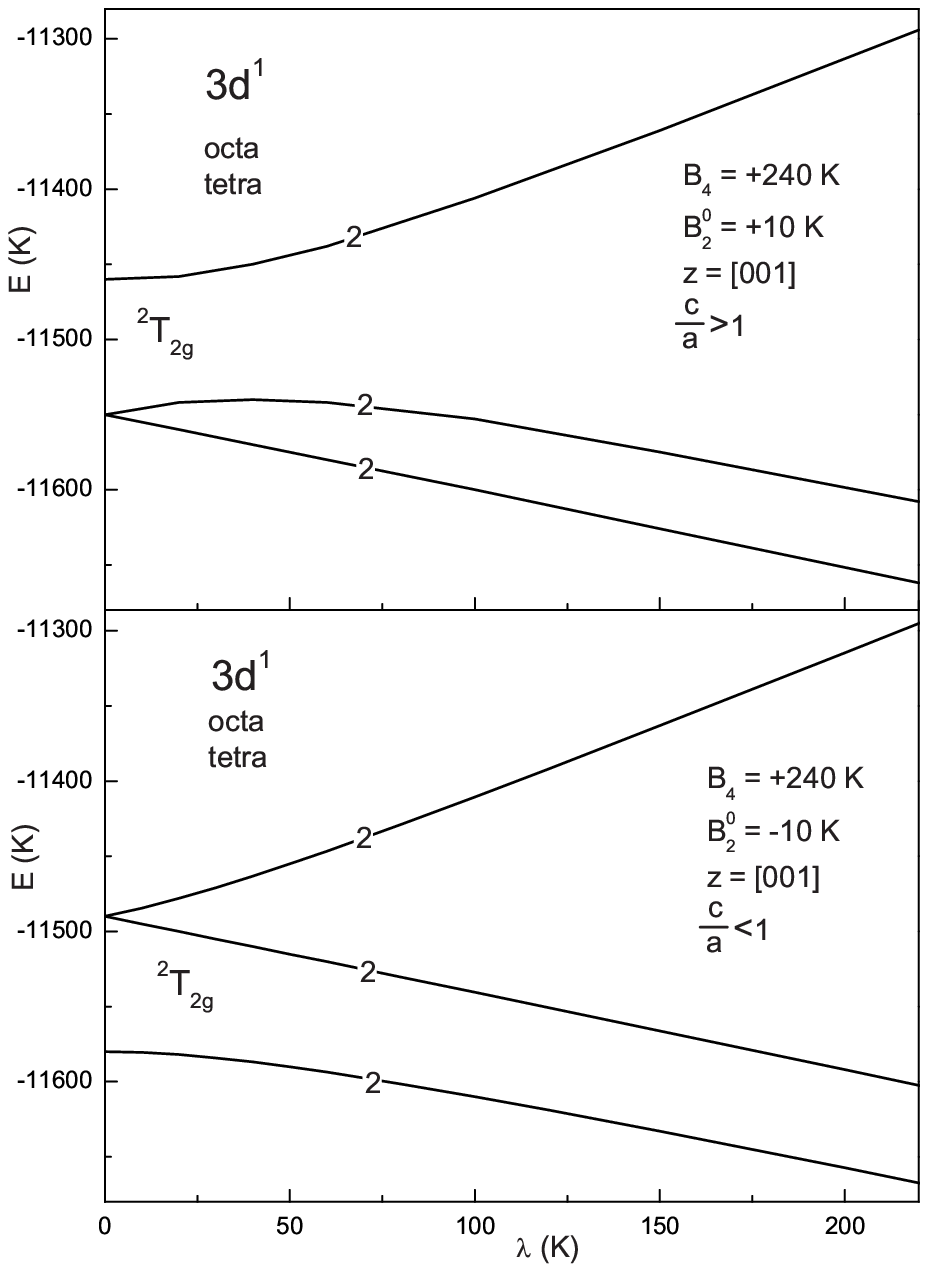}
\end{center} \vspace {-0.9 cm}
\caption{The calculated spin-orbit coupling $\lambda$ dependence
of the three lowest states $t_{2g}$ (=$^{2}T_{2g}$) of the
3$d^{1}$ configuration for the elongated (c/a$>$1) and stretched
(c/a$<$1) tetragonal off-octahedral distortion. The right hand
states correspond to the three lowest states shown in Fig. 1d and
1e.\vspace {-0.6 cm}}
\end{figure}

The detailed form of the Hamiltonian (1) is written down in the LS
space that is the 10 dimensional spin-orbital space $\left|
LSL_{z}S_{z}\right\rangle $. The $L$ and $S$ quantum numbers for
one 3d electron are equal to $L$=2 and $S$=1/2 (here, for 3$d^{1}$
configuration, lower $l$ and $s$ could be also used). The
Hamiltonian (1) is customarilly treated by perturbation methods
owing to the weakness of the s-o coupling for the 3$d$ ions in
comparison to the strength of the crystal-field interactions. We
have accepted the weakness of the s-o coupling, what is reflected
by the sequence of terms in the Hamiltonian (1), but we have
performed direct calculations treating all shown terms in the
Hamiltonian (1) on the same footing. The separate figures, if
presented, are shown for the illustration reasons.

Diagonalization have been performed for physically relevant values of $%
\lambda _{s-o}$ of +220 K (= 150 cm$^{-1}$) found for the
Ti$^{3+}$ ion \cite {21}. The cubic CEF parameter B$_4$ is taken
as +240 K (results are not sensitive to its exact value provided
B$_{4}$ $>$ +50 K). The positive sign of B$_4$ comes from {\it ab
initio} point charge calculations of octupolar interactions of the
Ti$^{3+}$ ion with the octahedral oxygen (negative charges)
surroundings. Such value of B$_4$ yields the $T_{2g}$-$E_{g}$
splitting of 2.5 eV. A splitting of 2.15-2.5 eV has been observed
for Ti$^{3+}$ ions in Al$_{2}$O$_{3}$, where the similar oxygen
octahedron exists. In LaCoO$_{3}$ we derived value of B$_{4}$ of
280-320 K \cite {26}.

In Fig. 2 we show detailed calculations of the influence of the
spin-orbit coupling on the three lowest states $t_{2g}$
(=$^{2}T_{2g}$) of the 3$d^{1}$ configuration for the elongated
(c/a$>$1) and stretched (c/a$<$1) tetragonal off-octahedral
distortion, realized for B$_2^0$ $>$ 0 and B$_2^0$ $<$ 0,
respectively. Analyzing the effect of the sign of the tetragonal
distortion lead us to a conclusion that the magnetic moment lies
along the tetragonal axis for the $z$-axis stretching. For the
elongation case the moment lies perpendicularly to the tetragonal
axis. Taking into account that in YTiO$_{3}$ the ordered moment
lies along the $c$ direction (in the $Pbnm$ structure) and our
long-lasting studies of CEF effects \cite {16,17,27,28} we came
to values for B$_2^0$ = -50 K and B$_2^2$ = -10 K which reproduce
the magnetic moment value, its direction and (some)
thermodynamics.
\begin{figure}[t]\vspace {-1.0 cm}
\begin{center}
\includegraphics[width = 7.7 cm]{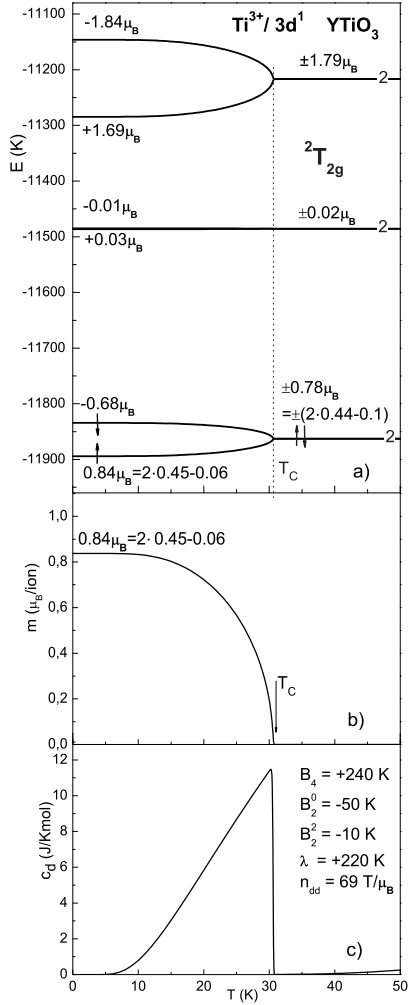}
\end{center}\vspace {-1.6 cm}
\caption{The calculated temperature dependence of some properties
of YTiO$_{3}$. a) the temperature dependence of the three lowest
states ($t_{2g}$ states) of the Ti$^{3+}$-ion in YTiO$_{3}$ in the
magnetically-ordered state below T$_{c}$ of 30.6 K; in the
paramagnetic state the electronic structure is temperature
independent unless we do consider a changing of the CEF
parameters, for instance, due to the thermal lattice expansion.
The used parameters: B$_4$= +240 K, B$_2^{0}$= -50 K, B$_2^{2}$=
-10 K, $\lambda _{s-o}$ = +220 K and n$_{d-d}$= 69 T/$\mu_{B}$.
The splitting of the Kramers doublets should be noticed in the
ferromagnetic state. Excited states are at 377, 645, 28843 and
29447 K. (b) the temperature dependence of the Ti$^{3+}$-ion
magnetic moment in YTiO$_{3}$. At 0 K the total moment m$_{Ti}$
of 0.84 $\mu _{B}$ is built up from the orbital m$_{o}$ and spin
m$_{s}$ moment of -0.06 and 0.90 $\mu _{B}$, respectively. c) The
calculated temperature dependence of the 3$d$ contribution
$c_{d}(T)$ to the heat capacity of YTiO$_{3}$. The $\lambda$-type
peak marks T$_{c}$.\vspace {-0.4 cm}}
\end{figure}
The calculated ground-state eigenfunction (the $z$ component of
$L$ and $S$ are shown)
\begin{center}
$\psi _{GS\pm}$ = 0.690$\left| \pm 2,\mp \frac 12\right\rangle $ -
0.678$\left| \mp 2,\mp \frac 12\right\rangle $ \\
- 0.253$\left| \pm 1,\pm \frac 12\right\rangle $ - 0.020$\left|
\mp 1,\pm \frac 12\right\rangle$ (2)
\end{center}

where the sign $\pm $ refers to two conjugate Kramers states.
This state has $S_{z}$ =$\mp $0.44 and $L_{z}$ =$\pm $ 0.10. The
resultant moment $m_{z}$ =$\pm $0.78 $\mu_{B}$ cancels each other
in the paramagnetic state as is denoted in Fig. 3a.

Making use of the $\left| xy,\mp \right\rangle $ function
extended for the spin component, as $\left| xy,\mp \right\rangle
$ $=$ $\sqrt{1/2\text{ }}\left( \left| 2,\mp \frac
12\right\rangle -\left| -2,\mp \frac 12\right\rangle \right) $
(also functions $\left| xz\right\rangle $, $\left| yz\right\rangle
$, $\left| x^{2}-y^{2}\right\rangle $ and $\left|
z^{2}\right\rangle $ ) one can write the ground state $\psi
_{GS\pm}$ function approximately as:
\begin{center}
$\psi _{GS\pm}$ = 0.967$\left| xy,\mp \right\rangle $ +
0.0085$\left| x^{2}-y^{2},\mp \right\rangle $ \\
- 0.186$\left| xz,\pm \right\rangle $ - 0.172$\left| yz,\pm
\right\rangle + ....(3)$
\end{center}

In the magnetic state the ground state $\psi _{GS\pm}$ Kramers
doublet function becomes polarized as a molecular field is
self-consistently settled down and the function
\begin{center}
$\psi _{GS+}$ = 0.695$\left| + 2,- \frac 12\right\rangle $ -
0.686$\left| - 2,- \frac 12\right\rangle $ \\
- 0.215$\left| + 1,+ \frac 12\right\rangle $ - 0.016$\left| - 1,+
\frac 12\right\rangle (4)$
\end{center}
is obtained as the ground state. The higher conjugate state is
calculated to be described by
\begin{center}
$\psi _{GS-}$ = 0.665$\left| + 2,+ \frac 12\right\rangle $ -
0.682$\left| - 2,+ \frac 12\right\rangle $ -
 \\
+ 0.303$\left| - 1,- \frac 12\right\rangle $ + 0.026$\left| + 1,-
\frac 12\right\rangle (5).$
\end{center}
\begin{figure}[t]\vspace {-1.0 cm}
\begin{center}
\includegraphics[width = 9.6 cm]{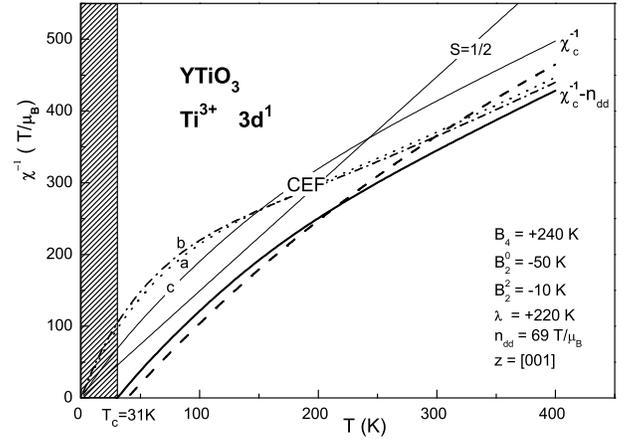}
\end{center}\vspace {-1.2 cm}
\caption{Calculated temperature dependence of the paramagnetic
susceptibility $\chi $ (T) for the 3d$^1$ system Ti$^{3+}$ in
YTiO$_{3}$ for three different crystallographic directions ($a$,
$b$, and $c$) calculated for B$_4$= +240 K, B$_2^{0}$= -50 K,
B$_2^{2}$= -10 K and $\lambda _{s-o}$ = +220 K; these curves are
denoted with CEF. The lowest solid line is the $\chi_{c} $ (T)
dependence calculated with taking into account the ferromagnetic
interactions with n$_{d-d}$= 69 T/$\mu_{B}$ - this curve should be
compared with experimental data. Experimental data, taking after
Ref. \cite {4}, are shown by the lowest dashed line. The shadow
area indicates the ferromagnetic state. The straight line denoted
with $S$ =1/2 shows the Curie law expected for the free $S$ =1/2
spin.\vspace {-0.4 cm}}
\end{figure}

Below T$_{c}$ there opens, as is seen in Fig. 3a, a spin-like gap
that amounts at T = 0 K to 59.7 K (= 41.4 cm$^{-1}$ = 5.15 meV).
The spin-like gap is associated with the splitting of the Kramers
doublet ground state in the ferromagnetic state. The magnetic
ground state $\psi _{GS+}$ has $S_{z}$ = -0.45 and $L_{z}$ =
+0.06 and the resultant moment amounts to 0.84 $\mu_{B}$. The
appearance of the magnetic state is calculated self-consistently.
It appears in the instability temperature (T$_{c}$) in the
temperature dependence of the CEF paramagnetic susceptibility when
\begin{center}
\vspace {-0.1cm} $\chi_{CF}^{-1}(T_{c})$ = $n_{d-d} $        (6)
\end{center}
as is illustrated in Fig. 4 for different crystallographic
directions. With decreasing temperature this equality is reached
the first for the $c$ direction pointing the preferred magnetic
arrangement axis. The magnetic state is calculated
self-consistently by adding to the Hamiltonian Eq. (1) the
inter-site magnetic (spin-dependent) interactions instead of the
last Zeeman term \cite {29} \vspace {-0.2 cm}
\begin{center}
$H_{d-d}$ = $n_{d-d}\left( -m_{d}\cdot m_{d} +
\frac{1}{2}\left\langle m_{d}^{2}\right\rangle \right) $ (7)
\end{center}
where $n_{d-d}$ is the molecular-field coefficient.

Having eigenvalues in a function of temperature we calculate the
free energy $F(T)$. From $F(T)$ using well-known statistical
formulae we calculate all thermodynamics like temperature
dependence of the magnetic moment, of the additional $c_{d}$ heat
capacity, of paramagnetic susceptibility, of the 3$d$-shell
quadrupolar moment and many others. The present calculations are
similar to those performed for FeBr$_{2}$ \cite {30} and CoO
\cite {31}. \vspace {-0.7cm}
\section {Conclusions} \vspace {-0.6 cm}We have calculated consistently a
value of the magnetic moment of 0.84 $\mu_{B}$ and its direction
(along the $c$ axis in the Pbnm structure) in YTiO$_{3}$ as well
as temperature dependence of the paramagnetic susceptibility in
very good agreement with experimental observations.

This remarkable reproduction of so many physical properties has
been obtained within the localized-electron approach taking into
account lattice off-octahedral distortions and the intra-atomic
spin-orbit coupling \cite {25}. Although the spin-orbit coupling
is weak, it amounts to only 1.2 \% of the octahedral CEF
splitting, it has enormous influence on the low-energy electronic
structure and low-temperature magnetic and electronic properties.
The spin-orbit coupling binds the orbital moment to the spin
moment being the reason for the spin-lattice and spin-phonon
coupling. The suggested energy positions of the excited CEF
states have to be verified by experiment. The good reproduction
of the paramagnetic susceptibility indicates on the proper overall
splitting of the $t_{2g}$ states - one should note that our
splitting is about 5 times smaller than that obtained in Refs
\cite {8,9,12,14}. We are aware that our calculations have to be
extended to take into account many other effects - first of them
seems to be geometrical effects associated with the
non-collinearity of the local symmetry axes, but these studies
prove that magnetic properties of YTiO$_{3}$ are predominantly
determined by the atomic-scale lattice distortions, crystal-field
and the spin-orbit coupling of the Ti$^{3+}$ ions, whereas charge
fluctuations are negligible. An interplay of the spin-orbit
coupling, lattice distortions and magnetic order is very subtle
involving rather small energies, smaller than 5 meV making
theoretical studies quite difficult.

We point out that all discussed by us parameters are physical
measurable parameters. The B$_{2}^{0}$ and B$_{2}^{2}$ parameters
are related to the observed orthorhombic local surroundings. A
negative value of the B$_{2}^{0}$ parameter results from the
stretching of the apical bonds with respect to the average bond
length within the $a-b$ plane. The B$_{2}^{2}$ parameter is
related to the difference in the $a-b$ plane. A success of our
ionic approach, called due to extension to the magnetic state a
quantum atomistic solid-state theory (QUASST) \cite {32}, if
applied to YTiO$_{3}$, is related to our long-lasting systematic
studies, despite of discrimination and inquisition, of the
spin-orbit coupling and crystal-field interactions in 3$d$
compounds and in study of the region, where the spin-orbit
coupling and off-octahedral lattice distortions are of the
comparable strength.

{\bf Acknowledgements}. We are very grateful to all our opponents.
Although we do not think that the discrimination and inquisition
should take place in Physics at the XXI century their critics
largely stimulated these, very natural for us studies. This
discrimination is the best proof that the knowledge about the CEF
and the role of the spin-orbit coupling in 3$d$ magnetism is
rather poor. We are thankful to a numerous members of the
International Committee of the Strongly-Correlated Electron
Systems Conference in Vienna 2005 for the friendly support. We
are convinced that Physics can develop only in the friendly
atmosphere and in the open exchange of information.

$^\spadesuit$ dedicated to John Van Vleck and Hans Bethe, pioneers
of the crystal-field theory, to the 75$^{th}$ anniversary of the
crystal-field theory, and to Pope John Paul II, a man of freedom
and honesty in life and in Science.


\begin{thebibliography}{9}
\bibitem{25} the computer program is available
on the written request to the authors.
\bibitem{29} the use of the spin-spin Hamiltonian lead to similar
results - we checked it getting n$_{ex}$= 63.75 T/$\mu_{B}$ in
accordance with the scaling of n$_{d-d}$ with the spin/total
moment ratio.
\end{thebibliography}
\end{document}